\documentclass[prl,aps,twocolumn,showpacs]{revtex4}

\usepackage{graphicx}

\begin{document}

\title{Electronic, Mechanical, and Piezoelectric Properties of ZnO Nanowires}

\author{H. J. Xiang}
\affiliation{Hefei National Laboratory for Physical Sciences at
  Microscale,
  University of Science and Technology of
  China, Hefei, Anhui 230026, People's Republic of China}

\author{Jinlong Yang}
\thanks{Corresponding author. E-mail: jlyang@ustc.edu.cn}

\affiliation{Hefei National Laboratory for Physical Sciences at
  Microscale,
  University of Science and Technology of
  China, Hefei, Anhui 230026, People's Republic of China}

\author{J. G. Hou}
\affiliation{Hefei National Laboratory for Physical Sciences at
  Microscale,
  University of Science and Technology of
  China, Hefei, Anhui 230026, People's Republic of China}

\author{Qingshi Zhu}
\affiliation{Hefei National Laboratory for Physical Sciences at
  Microscale,
  University of Science and Technology of
  China, Hefei, Anhui 230026, People's Republic of China}

\date{\today}

\begin{abstract}
  Hexagonal [0001] nonpassivated ZnO nanowires are studied
  with density functional calculations.
  The band gap and Young's modulus in nanowires which
  are larger than those in bulk ZnO
  increase along with the decrease of the radius of nanowires.
  We find ZnO nanowires have larger effective piezoelectric constant 
  than bulk  ZnO due to their free boundary. In addition, the effective piezoelectric
  constant in small ZnO nanowires doesn't depend monotonously on the radius
  due to two competitive effects: elongation of the
  nanowires and increase of the ratio of surface atoms.
\end{abstract}


\maketitle

ZnO\cite{Ozgur2005} is one of the most important materials due to its three key 
advantages: semiconducting 
with a direct wide band gap of 3.37 eV and a large
excitation binding energy (60 meV), piezoelectric  
due to non-central symmetry in the wurtzite structure, and
biocompatible. 
Recently, a diversity group of ZnO nanostructures 
including nanowires\cite{Huang2001}, nanobelts\cite{Pan2001}, 
nanosprings\cite{Kong2003}, 
nanocombs\cite{Wang2003},
nanorings\cite{Kong2004}, 
nanobows\cite{Hughes2004}, 
and nanohelices\cite{Yang2004,Gao2005} have
been synthesized under specific growth conditions.
ZnO nanostructures could have novel applications due to their 
unique physical and chemical properties arising from surface and
quantum confinement.
In particular, ZnO nanowires with relatively simple structures
are important one-dimensional (1D) nanostructures.
Experimentally, the group of Wang had synthesized well-aligned [0001]
ZnO nanowires enclosed by facet \{10$\bar{1}$0\} surfaces
\cite{ZnO_nanowire1,ZnO_nanowire2}.
Room-temperature ultraviolet lasing\cite{ZnO_UV}
and piezoelectric nanogenerators based on ZnO nanowire arrays
have been demonstrated\cite{Wang2006_science}. 
Rectifying diodes of single ZnO nanobelt/nanowire-based devices \cite{Lao2006} and
a ZnO nanowire photodetector\cite{Law2006}  were
fabricated very recently.

Although many studies on ZnO nanowires have been
conducted, there are some important issues remained to be addressed. 
First, the mechanical properties, especially the Young's
modulus of ZnO nanowires are on debate 
in the literature\cite{Bai2003,Yum2004,Song2005,
  Kulkarni2005,Chen2006}. For instance, Chen {\it et al.}
\cite{Chen2006} showed that the Young' modulus of ZnO nanowire with
diameters smaller than about 120 nm is significantly higher than 
that of bulk ZnO. However, the elastic modulus of vertically aligned
[0001] ZnO nanowires with an average diameter of 45 nm measured by
atomic force microscopy was found to be far smaller than that of bulk
ZnO\cite{Bai2003}. 
The second issue is about the electromechanical coupling in ZnO nanowires.
The effective piezoelectric coefficient of individual (0001) surface
dominated ZnO 
nanobelts measured by piezoresponse force microscopy was found to be much
larger than the value for bulk wurtzite ZnO\cite{ZnO_PFM}.  
In contrast, Fan {\it et al.} showed that the piezoelectric
coefficient for ZnO nanopillar with the 
diameter about 300 nm is smaller than the bulk
values\cite{Fan2006}. They suggested that the reduced
electromechanical response might be due to structural defects in the
pillars\cite{Fan2006}. Whether the electromechanical coupling is
enhanced or depressed in defect-free ZnO nanowires is not clear.  
Thirdly, although it is well known that the quantum confinement effect will
decrease the band gap of passivated nanowires, the question that how 
the dangling bond in bare ZnO nanowires affects the band gap
remains open.
The fundamental study on these issues is crucial for
developing future applications of ZnO nanowires.

In this letter, we have studied the electronic, mechanical,
and piezoelectric properties of [0001] ZnO nanowires using
first-principles methods for the first time.
We find that the band gap increases along with the decrease of
the radius of ZnO nanowires due to the radial confinement. 
The Young's modulus of nanowires is larger than bulk ZnO, in agreement
with the experimental results of Chen {\it et al.}\cite{Chen2006}.
The effective piezoelectric constant in ZnO nanowires is larger than
that of bulk ZnO due to the free boundary of nanowires.
Moreover, the effective piezoelectric constant in small ZnO nanowires doesn't
depend monotonously on the radius due to two competitive effects.

Our calculations are performed using the SIESTA package\cite{siesta}, a
standard
Kohn-Sham density-functional program using norm-conserving
pseudopotentials and numerical atomic orbitals as basis sets.
The local density approximation (LDA)\cite{LDA} to the exchange 
correlation functional is employed. 
The DZP basis sets for both Zn and O with the energy shift parameter
20 mRy are used.
The mesh cutoff parameter for real space integration is set to 350 Ry
to obtain accurate atomic forces.
Bulk ZnO has a wurtzite structure with the noncentral symmetry,
resulting in a normal dipole moment
and spontaneous polarization along the $c$-axis. 
The computed (experimental\cite{ZnO_exp}) lattice parameters of
bulk wurtzite ZnO are
$a=3.17$ \AA (3.25 \AA), $c=5.18$
(5.20 \AA), $u=0.374$ (0.381). 
The calculated LDA direct band gap is 0.63 eV,
in good agreement with
other LDA results\cite{DFPT_Vanderbilt,Usuda2002}. 

\begin{figure}[!hbp]
  \includegraphics[width=7.5cm]{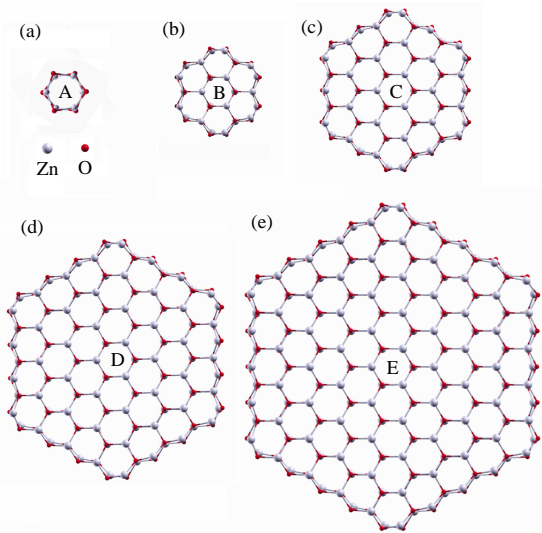}
  \caption{(Color online) Relaxed structures for ZnO nanowires with
  different radius. We label these nanowires as A, B, C, D, and E.}
  \label{fig1}
\end{figure}

\begin{table}
  \caption{Unrelaxed diameter
    ($D_0$) relaxed diameter ($D$), relaxed lattice constant ($c$), strain energy
    ($E$), band gap ($E_g$), Young's modulus ($E_{3}$), 
    and effective piezoelectric constant ($e_{33}^a$)
    of five different ZnO
    nanowires ( A, B, C, D, and E) and bulk ZnO. }
  \begin{tabular}{ccccccc}
    \hline
    \hline
    & A  &B & C & D & E & bulk \\
    \hline
    $D_0$ (\AA) & 3.66 & 9.68 & 15.96 & 22.27  & 28.59  &  \\
    $D$ (\AA)  & 3.32 & 9.32 &  15.61 &  21.97 &  28.33 & \\ 
    $c$ (\AA) & 5.335 & 5.302 & 5.270 & 5.234 & 5.215 & 5.180 \\
    $E_g$ (eV) & 2.40 & 1.54 & 1.09 & 0.85 & 0.75 & 0.63 \\
    $E_{3}$ (GPa) & 363 & 242 & 217 & 189 & 182 & 147 \\
    $e_{33}^a$ ($10^{-16} \mu C $\AA/ion) & 2025 & 1837 & 1879 & 1986
    & 1961 & 1453 \\
    \hline
    \hline
  \end{tabular}
  \label{table1}
\end{table}

Here we mainly focus on [0001] ZnO
nanowires since almost all ZnO nanowires grown along the [0001] 
direction\cite{ZnO_nanowire1,ZnO_nanowire2}. 
The nanowires are enclosed by six facet \{10$\bar{1}$0\} surfaces with low surface energy.
In wurtzite bulk ZnO, each Zn (O) bonds with four O (Zn) atoms.
In nanowires, the surface Zn (O) atom is only bound to three nearest
neighbours.   
In this study, we consider only bare ZnO nanowires without 
saturating the surface atoms.  
The nanowires are modeled by hexagonal supercells
whose lateral lattice constants are so large that 
there is almost no interaction between the nanowires.
Five ZnO nanowires with diameter (Here the diameter is defined as the
largest lateral distance between atoms)
ranging from about 0.3 to 2.8 nm, labeled as A, B,
C, D, and E respectively, are examined.
The largest nanowire E contains 300 atoms in the unitcell.
Since the lattice constant of small ZnO nanowires might differ
significantly from the bulk counterpart, we optimize both the lattice
constant $c$ and internal coordinates of nanowires.
The relaxed structures of these ZnO nanowires are shown in
Fig.~\ref{fig1}.
Since both surface Zn and O atoms move inwards and Zn
atoms move much more, it looks like surface O atoms rotate outwards.
The diameters of relaxed and unrelaxed nanowires are shown in
Table~\ref{table1}. We can see that the diameters of relaxed nanowires  
are smaller than those of unrelaxed nanowires by almost 0.3 \AA.
The relaxation of surface atoms in ZnO nanowires is similar to that in
ZnO [10$\bar{1}$0] surface\cite{Schroer1994}. 
Along with the shrinkage of surface atoms, 
the lattice contant $c$ of ZnO nanowires is increased when compared
with that of bulk ZnO, as are shown in
Table~\ref{table1}. We can see that the elongation of small nanowires
is considerably large, however, the lattice constants of large
nanowires tend to approach that of bulk ZnO.

\begin{figure}[!hbp]
  \includegraphics[width=7.5cm]{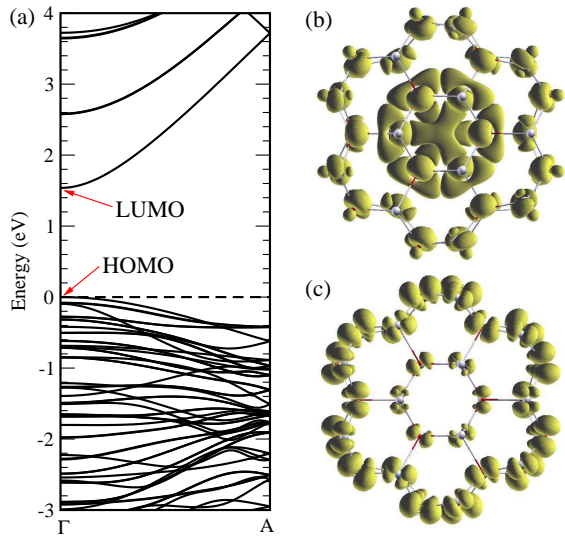}
  \caption{(Color online) (a) The band structure of nanowire B.
    (b) The charge density of the LUMO state at $\Gamma$ for nanowire
    B. (c) The charge density of the HOMO state at $\Gamma$ for
    nanowire B.}
  \label{fig2}
\end{figure}

We have calculated the electronic structures of these ZnO nanowires.
All [0001] ZnO nanowires are found to be semiconducting. 
Although the LDA usually underestimates the band gap,   
the trend of the band gaps of ZnO nanowires predicted from the LDA
calculations are expected to be correct.
The LDA band gaps are shown in Table~\ref{table1}. Clearly, 
the band gaps of ZnO nanowires
increase monotonously along with the decrease of the radius of
nanowires. In comparison with bulk ZnO, the increment of the band gap
of nanowire A can be as large as $1.77$ eV.
The blueshift of the band gap should be due to the quantum confinement
effect. However, the gap-broadening effect in nonpassivated nanowires  
is unusual. For example, nonpassivated Si nanowires grown along the
$\langle 100 \rangle$ direction are found to be metallic and semimetallic due to the
presence of surface states\cite{Rurali2005}. To see the details of the
electronic structure of ZnO nanowires, we plot the band structure of
nanowire B in Fig.~\ref{fig2}(a). The band structures of other ZnO
nanowires are similar to that of nanowire B. Clearly, ZnO nanowires
have a direct gap at $\Gamma$. To gain an insight into  the character of the 
lowest unoccupied molecular
orbital (LUMO) and highest
occupied molecular orbital (HOMO), we show the charge density of the
LUMO and HOMO states at $\Gamma$ in Fig.~\ref{fig2}(b) and (c),
respectively. 
We can see that the LUMO is delocalized in the whole nanowire,
indicating that it is a bulk state instead of a surface state. 
The delocalized distribution is also responsible for the   
large dispersion of the LUMO from $\Gamma$ to $A$. Furthermore, due to
the delocalized character, the LUMO energy will increase substantially
in small radius nanowires due to the radial confinement. 
Since the charge density of the HOMO at $\Gamma$ mainly contributed by
surface atoms, the HOMO is a surface state. The HOMO at $\Gamma$ lies
only 80 meV above the top of valence band of bulk ZnO, and its
position changes little in nanowires with different diameters 
since the HOMO is mainly composed by surface O 2p like dangling
bonds. The HOMO is similar to the P1 surface state in the ZnO
(10$\bar{1}$0) surface as identified by Schr\"oer {\it et al.}\cite{Schroer1994}.
It is the different response to the radial confinement between the
LUMO and HOMO that leads to the overall gap-broadening effect in ZnO
nanowires.

Now we turn to study the mechanical properties of ZnO nanowires. 
An important mechanical parameter for describing a one-dimensional
system is the Young's modulus. We calculate $E_{3}$ in the [0001] direction by using the
following formula:
\begin{equation}
  E_{3}=\frac{1}{V} d^{2} E / d \epsilon ^{2}, 
\end{equation}
where $E$ is the total energy, $\epsilon$ is the axial strain, and $V$
is the system volume. 
For nanowires, we define the volume as $V=S\times c$, where $S$ is the area of
the cross section of the nanowire and $c$ is the lattice constant.
For each nanowire, we perform seven calculations with
$\epsilon$ range from $-3$\% to $3$\%. In each calculations for the
nanowires, we fully relaxed the internal coordinates. To compare our
results for nanowires with bulk ZnO, we also calculate the Young's
modulus of bulk ZnO. In the calculations, only the lattice constant
$c$ is fixed to the strained value, all other degrees of freedom
including the lateral lattice constants are optimized. Our results for
bulk ZnO and ZnO nanowires are reported in Table~\ref{table1}. 
The calculated Young's modulus of bulk ZnO is 147 GPa, which is
close to the value 140 GPa deduced from the experimental elastic
constants\cite{Kobiakov1980}. For ZnO nanowires, the Young's modulus
increase monotonously along with the decrease of the radius of
nanowires and are larger than that of bulk ZnO.
To estimate the Young's modulus of larger ZnO nanowires, the
calculated results are fitted using $E_3(D)=145.87+1172.63 D^{-1}
-1477.58 D^{-2} + 4830.55 D^{-3}$, where Young's modulus $E_3$ is in unit of
GPa and diameter $D$ is in unit of \AA.
The trend of Young's modulus accords qualitatively with the experimental
results\cite{Chen2006}.
Our results also
agree qualitatively with the Young's modulus of ZnO nanobelts
with rectangular cross-sections in the [0001] orientation calculated
from empirical molecular dynamics simulations\cite{Kulkarni2005}. 
However, the experimental Young's modulus of ZnO nanowires are 
significantly larger than our results. For example, the Young's
modulus of a ZnO nanowire with the diameter 20 nm estimated using their fitted
formula is 202 Gpa, however, our value estimated from the fitted formula
is only 152 Gpa. The reason for the discrepancy is not very
clear. 

Before we proceed to discuss the piezoelectric effect in ZnO
nanowires, we first calculate piezoelectric constant $e_{33}$ of bulk
ZnO. 
The piezoelectric constants $e_{ij}$ is defined as follows:
$e_{ij}=\partial{\mathbf{P}_{i}}/\partial{\epsilon_{j}}$, where
$\mathbf{P}$ is the total polarization, and $\epsilon_{j}$ is the
strain tensor component. 
Here, the piezoelectric constants are
computed by combining the Berry
phase method in the modern theory of polarization\cite{mod_polarization} with the finite
difference method.
The piezoelectric effect in tetrahedrally bonded
semiconductors results from two different terms of opposite sign: The
``clamped-ion'' and the ``internal-strain'' contributions.
The calculated clamped-ion and relaxed-ion piezoelectric constant $e_{33}$ for
bulk ZnO is $-0.77$ and $1.29$ C/m$^2$ respectively,
which accords with others' results ($-0.75$ and $1.28$
C/m$^2$)\cite{DFPT_Vanderbilt}.

\begin{figure}[!hbp]
  \includegraphics[width=8.5cm]{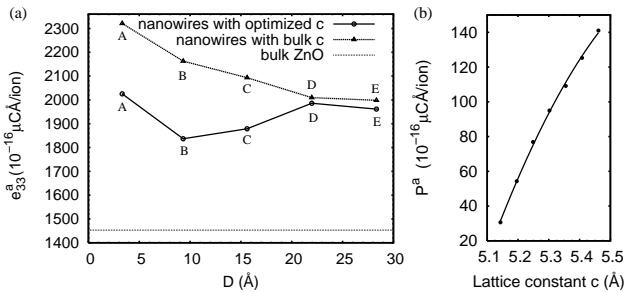}
  \caption{(a) The dependence of the effective piezoelectric constant
    $e_{33}^a$ upon the diameters of the nanowires
    with optimized lattice constant $c$ and nanowires with 
    bulk lattice constant $c$. 
    The horizontal dotted line indicates the piezoelectric constant of bulk ZnO.
    (b) shows the polarization of ZnO nanowire B with different $c$.}
  \label{fig3}
\end{figure}

As the conventional definition of piezoelectric constant
for three-dimensional bulk is not appropriate for describing the
piezoelectric properties of one-dimensional systems, we define the
atomic averaged effective piezoelectric constant as: $e_{33}^a$=$e_{33}\times V_{\textrm{scell}}
/N$, where $N$ is the number of atoms, and $V_{\textrm{scell}}$ is the
volume of the supercell. 
The results are presented in Fig.~\ref{fig3}(a) and
Table~\ref{table1}. 
We can clearly see that $e_{33}^a$ of nanowires is considerably larger than that
of bulk ZnO. 
We find that larger $e_{33}^a$ in nanowires is caused
by the free boundary of nanowires. 
In the calculations of $e_{33}^a$, the lateral lattice
constants are fixed. When bulk ZnO is strained along the $c$ axis,
atoms can not relaxed freely along the lateral directions.
However, when nanowires are compressed or
elongated along the $c$ axis, atoms can relaxed freely along the
lateral directions due to the free boundary. Hence, the effective
piezoelectric constant $e_{33}^a$ in nanowires should approach $e_{33}^{b}
- 2 e_{31}^{b} \times  \nu $, where $e_{33}^{b}$
and $e_{31}^{b}$ is effective piezoelectric constants of bulk ZnO, 
$\nu=-\epsilon_1/\epsilon_3=-\epsilon_2/\epsilon_3$ is Poisson's ratio
($\epsilon_1=\epsilon_2=(a-a_0)/a_0$, $\epsilon_3=(c-c_0)/c_0$, $a$ is
the relaxed lateral lattice constant when the lattice constant of ZnO
is changed to $c$). Since $\nu > 0$ and $e_{31}^{b} <
0$\cite{DFPT_Vanderbilt}, $e_{33}^a$ in nanowires is larger than $e_{33}^{b}$. 
For the same reason, effective piezoelectric coefficient in ZnO
nanobelts was also found experimentally to be larger than that of bulk ZnO by Zhao 
{\it et al.}\cite{ZnO_PFM}.
One might expect smaller nanowires have larger piezoelectric
constant due to their larger ratio of surface atoms.
However, Fig.~\ref{fig3}(a) shows that $e_{33}^a$
doesn't monotonously depend on the radius of ZnO nanowires:
$e_{33}^a$ for nanowire D is larger than that for nanowire C and E.
We attribute this anormal behavior
to the increase of the lattice constant
along with the decrease of the radius of ZnO nanowires, 
since the piezoelectric constant of a certain nanowire decreases
along with the increase of the lattice constant $c$, as shown in 
Fig.~\ref{fig3}(b). 
In fact, we also calculate the effective piezoelectric constant of the
nanowires with bulk lattice constant $c$, and the results (also shown
in Fig.~\ref{fig3}(a)) indicate a monotonous decreasing
dependence of $e_{33}^a$ upon the radius of nanowires. From Fig.~\ref{fig3}(a), we
can also see that the difference of $e_{33}^a$ between nanowires D and
E with the lattice constant $c$ relaxed or fixed is very small,
suggesting that larger nanowires will have similar $e_{33}^a$ as nanowire E. 

To summarize, we have carried out comprehensive first-principles
studies on [0001] ZnO nanowires.
The band gaps of ZnO nanowires
increase along with the decrease of the radius of nanowires due to
the quantum confinement effect.
The Young's modulus of thiner nanowires is larger than that of
thicker  nanowires. 
Our calculations indicate that the effective piezoelectric constant
$e_{33}^a$ of ZnO nanowires is larger than that of bulk ZnO. In
addition, we find a non-trivial dependence of the electromechanical 
coupling of ZnO nanowires upon the radius as a result of the
competition between two opposite factors. 
Our results support the application of ZnO nanowires as
nanosensors and nanoactuators.

This work is partially supported by the National Natural Science
Foundation of China (50121202, 20533030, 10474087), by the USTC-HP HPC
project, and by the SCCAS and Shanghai Supercomputer Center.


\end{document}